\documentstyle{elsart} 
\newcommand{\Df}[2]{\mbox{$\frac{#1}{#2}$}}

\begin{document}

{\small \hfill  INP-98-7/508, MPI/PhT/98-19, hep-ph/9802429, February
1998}

\vspace{2.5em}

\centerline{\Large\bf Three-loop vacuum integrals in FORM and REDUCE}

\vspace{2.5em}

\centerline{P.\,A.\,BAIKOV\footnote{
Supported in part by 
INTAS (grant 93--0744--ext), Volkswagen Foundation (contract
No.~I/73611)\\
\hspace*{3pt}Email: baikov@theory.npi.msu.su}
}
\centerline{\it Institute of Nuclear Physics, Moscow State University,}
\centerline{\it 119~899, Moscow, Russia}

\vspace{1.5em}

\centerline{M.\,STEINHAUSER}
\centerline{\it Max-Planck-Institut f\"ur Physik,Werner-Heisenberg-Institut,}
\centerline{\it D-80805 Munich, Germany}

\vspace{2.5em}

\centerline{\bf Abstract}

\begin{center}
\begin{minipage}{14cm}
The implementation of an algorithm for three-loop massive vacuum
integrals, based on the explicit solution of
the recurrence relations, in REDUCE and FORM is described.
\end{minipage}
\end{center}

\section{Introduction}

The increasing experimental precision makes it mandatory to compute
higher order corrections to physical quantities.
This essentially requires the evaluation of multi-loop integrals.
However, even at two loops it is very often not possible to
solve the integrals exactly and one has to rely on approximations.
A very powerful approach to get, nevertheless, reasonable
results is based on expansions in small quantities. 
Recently a method was developed for three-loop 
polarization functions combining expansions
from different kinematical regions with the help of conformal 
mapping and Pad\'e approximation~\cite{BaiBro95,chetn}.
An essential ingredient for this procedure are massive integrals
with vanishing external momentum. For their computation usually
recurrence relations are used which are based on the integration-by-parts
technique~\cite{ch-tk,REC}.
These relations connect Feynman integrals with different powers
of their denominators. In many cases
they provide a possibility to express an
integral with given degrees of the denominators as a linear
combination of a few so-called master integrals with prefactors 
which are rational functions of the space-time dimension $D$.
The construction of such a procedure is a nontrivial problem
even at two-loop level~\cite{Tar97}. At three loops up to now
only the case of vacuum integrals with one non-zero mass 
and various numbers of massless
lines has been considered~\cite{REC,Avd}.
For the problems of practical interest the direct application
of these equations usually leads to intermediate expressions which
need several hundred megabyte up to a few gigabyte of disk space.

For the two-loop massless master integral some years ago
the recurrence procedure has been solved explicitly~\cite{Tka83}.
The solution, expressed in terms of multiple sums, is also
used in practice~\cite{mincer} --- at least for those cases
where the exponents of the propagators are not too small.
Concerning the general multi-loop case
in~\cite{ES} a new approach to implement recurrence 
relations~\cite{ch-tk} was suggested.
There the factors in front of the
master integrals, which in the following are called coefficient functions,
are considered as independent solutions of the recurrence relations. 
In \cite{ES} integral representations for these solutions
were obtained. It appeared, that in some cases these representations can
be expressed in terms of Pochhammer symbols.
As an example the vacuum integrals with four equal masses and two
massless lines has been considered, and the efficiency
of this approach was demonstrated by the calculation of the three-loop 
QED vacuum polarization.

In this paper we describe the algorithm, suggested in~\cite{ES}, in
more details and discuss some peculiarities of its implementation in 
REDUCE~\cite{REDUCE}
and 
FORM~\cite{FORM}.
In Section~\ref{secfor} the general $L$-loop case is considered.
In Section~\ref{secthr} the derived formulas are specified to
three loops and explicit solutions are given. The implementation is discussed
in Section~\ref{secimp} and finally Section~\ref{seccon} contains
our conclusions.

\section{\label{secfor}Basic formulas}

Let us in a first step derive the recurrence relations.
The combinatoric structure of these relations becomes more
transparent if we start with the general multi-loop case, keeping in 
mind, however, the application at three-loop level.
Therefore let us consider $L$-loop vacuum integrals 
with $N=L(L+1)/2$ denominators
(This number of denominators provides the possibility to express any
scalar product of loop momenta as linear combination of the denominators;
the diagrams of practical interest which usually have less number of 
denominators, can be considered as special cases with some exponents 
equal to zero.):

\begin{eqnarray}
B(\underline{n},D)\equiv
B(n_1,\ldots,n_N,D)=
\Df{m^{2\Sigma n_i-LD}}
{\big[\imath\pi^{D/2}\Gamma(3-D/2)\big]^L}
\int \cdots \int \Df{d^Dp_1\ldots d^Dp_L} 
{D_1^{n_1}\ldots D_N^{n_N}},
\label{eqbn}
\end{eqnarray}
where $p_i$ (${i=1,\ldots,L}$) are loop momenta 
and $D_a=
A^{ij}_a p_i\cdot p_j -\mu_a m^2$ (a=1,\ldots,N; here and below
the sum over repeated indices is understood).
For convenience we set $m=1$ in the following.
The recurrence relations are obtained 
by acting with $(\partial/\partial p_i)\cdot p_k$ on
the integrand~\cite{ch-tk}:

\begin{eqnarray}
D\delta_k^i B(\underline{n},D)&=&
2
\tilde{A}_{kl}^a({\bf I}^-_a+\mu_a) 
A_d^{il}{\bf I}^{d+} 
B(\underline{n},D),
\label{rr}
\end{eqnarray}

\noindent
where  
${\bf I}^-_c B(\ldots, n_c,\ldots )\equiv B(\ldots, n_c-1,\ldots )$
and
${\bf I}^+_c B(\ldots, n_c,\ldots )\equiv n_c B(\ldots, n_c+1,\ldots)$
(here, no sum over $n_c$ may be performed).
The factors $\tilde{A}_{kl}^a$ arise from the fact that the 
scalar products $p_i\cdot p_j$ in the numerator which appear as 
a result of the differentiation are expressed in terms of the denominators
$D_a$ via: 
\begin{eqnarray}
p_k\cdot p_l&=&
\tilde{A}_{kl}^a(D_a+\mu_a)
\,.
\nonumber
\end{eqnarray}

Using the identities
\begin{eqnarray}
[{\bf I}^-_a,{\bf I}^{d+}]=\delta_a^d,
&&\qquad
A_a^{il} \tilde{A}_{kj}^a=
\Df{1}{2}(\delta^i_k\delta^l_j+\delta^i_j\delta^l_k)
\,,
\nonumber
\end{eqnarray}
the recurrence relations (\ref{rr}) can be represented as
\begin{eqnarray}
(D-L-1)\delta_k^i B(\underline{n},D)&=&2
A_d^{il}{\bf I}^{d+} 
\tilde{A}_{kl}^a({\bf I}^-_a+\mu_a) 
B(\underline{n},D),
\label{rr1}
\end{eqnarray}
where we have exploited that the matrices $A^a$ and $\tilde{A}^a$ can be 
chosen to be symmetrical.
If we denote
\begin{eqnarray}
\tilde{A}_{kl}^a({\bf I}^-_a+\mu_a)\equiv {\bf A}_{kl}, 
\qquad A_d^{il}{\bf I}^{d+}\equiv \mbox{\boldmath $\partial$}^{il}
\,,
\nonumber
\end{eqnarray}
Eq.~(\ref{rr1}) will read:
\begin{eqnarray}
\left[
\mbox{\boldmath $\partial$}^{il}\cdot{\bf A}_{kl}
-\Df{D-L-1}{2}\delta_k^i
\right]
B(\underline{n},D)&=&0.
\label{eq1}
\end{eqnarray}

Let us now diagonalize these relations with respect to the
operators $\mbox{\boldmath $\partial$}^{il}$. Therefore we multiply
Eq.~(\ref{eq1}) with $\mbox{$\partial$}^{jk} \det({\bf A})$,
which is the cofactor of the matrix element 
${\bf A}_{kl}$, and sum afterwards 
over $k$. Finally we arrive at:
\begin{eqnarray}
\left[
\mbox{\boldmath $\partial$}^{il}\cdot
\det({\bf A})
-\Df{D-L-1}{2}(\mbox{$\partial$}^{il}\det({\bf A}))
\right]
B(\underline{n},D)&=&0
\,.
\nonumber
\end{eqnarray}

Our goal is to find the coefficient functions $f^k(\underline{n},D)$
which relate the given integral $B(\underline{n},D)$ with the basic master
integrals $B(\underline{n}_{k},D)$:
\begin{eqnarray}
B(\underline{n},D)&=&
f^k(\underline{n},D)B(\underline{n}_{k},D)
\,,
\nonumber
\end{eqnarray}
where the coefficient functions have to fulfill the 
initial conditions $f^i(\underline{n}_k,D)=\delta^i_k$.
The index $k$ is used to label different sets of indices $\underline{n}$
which are fixed by the choice of the master integrals.
Assuming that the master integrals are algebraically independent
one can conclude that the
functions $f^k(\underline{n},D)$ should be independent solutions of the
recurrence relations. This means that as soon as we find a set of independent 
solutions it is possible to construct the desired coefficient 
functions as linear combinations respecting the given initial conditions. 
To this end it turns out that it is useful to construct an auxiliary
integral representation for these functions where the operators
can be written in the form\footnote{
  Note that before the transition to this representation is performed
  the order of the operators has to be reversed.}:
${\bf I}^{d+} \rightarrow \partial/\partial x_d$,
${\bf I}^-_d \rightarrow x_d$.
Then the differential equation
corresponding to Eq.~(\ref{rr1}) has the solution 
$g(x_a)=\det({\bf A})^{(D-L-1)/2}=P(x_a+\mu_a)^{(D-L-1)/2}$, 
where
$P(x_a)$ is a polynomial in $x_a$ of degree $L$:
\begin{eqnarray}
P(x_a)&=&\det(
\tilde{A}_{kl}^a x_a)
\,.
\nonumber
\end{eqnarray}
Therefore it is tempting to consider the ``Laurent''
coefficients of the function $g(x_a)$ in order to 
determine the coefficient functions:
\begin{eqnarray}
f^k(\underline{n},D)&=&
\Df{1}{(2\pi\imath)^N}
\oint \cdots \oint
\Df
{dx_1 \cdots dx_N}
{x_1^{n_1} \cdots x_N^{n_N}}
\det(\tilde{A}_{il}^a(x_a+\mu_a))^{(D-L-1)/2}
\,.
\label{solution}
\end{eqnarray}
Here the integral symbols denote $N$ subsequent complex 
integrations. The contours depend on the index $k$ and will be specified
below. Acting with Eq.~(\ref{rr1}) on~(\ref{solution}) 
leads (up to surface terms) to 
the corresponding differential operator acting on 
$g(x_a)$, which gives zero.
The surface terms can be removed by a proper choice of the
complex contours: either closed contours or 
integration paths which end in infinity.
For the last case one has to consider  
analytical continuations of $D$ from large negative values.
Note that Eq.~(\ref{solution}) is a solution of relation~(\ref{rr})
and thus the different choices of the contours correspond to
different solutions which are enumerated with the index $k$. 

The solutions (\ref{solution}) satisfy 
by construction the following condition:
\begin{eqnarray}
f^k(\underline{n},D)&=&P({\bf I}^-_a+\mu_a)f^k(\underline{n},D-2)
\,.
\label{rrD0}
\end{eqnarray}
In general, however, we are interested in 
coefficient functions, $\tilde{f}^k(\underline{n},D)$, 
which correspond to a specific set of
master integrals and hence a linear combination of
$f^k(\underline{n},D)$ with coefficients depending on
$D$ has to be considered. 
Then Eq.~(\ref{rrD0}) gets more complicated
as a mixing among the different functions is possible:
\begin{eqnarray}
\tilde{f}^k(\underline{n},D)&=&
S^k_i(D)P({\bf I}^-_a+\mu_a)\tilde{f}^i(\underline{n},D-2),
\nonumber
\end{eqnarray}
where $S_i^k(D)$ are rational functions of the space-time dimension $D$ 
only.

\section{\label{secthr}Three-loop case}

In this section we specify the general $L$-loop formulas derived in the
previous one to the three-loop case
with four equal masses
and two massless lines, i.e., $\mu_1=\mu_2=0,\mu_3=\mu_4=\mu_5=\mu_6=1$.
Then Eq.~(\ref{eqbn}) becomes:
\begin{eqnarray}
B(\underline{n},D)=
\Df{m^{2\Sigma_1^6 n_i-3D}}
{\big[\imath\pi^{D/2}\Gamma(3-D/2)\big]^3}
\int\int\int \Df{d^Dp\,d^Dk\,d^Dl} 
{D_1^{n_1}D_2^{n_2}D_3^{n_3}D_4^{n_4}D_5^{n_5}D_6^{n_6}}
\,,
\nonumber
\end{eqnarray}
with
\begin{eqnarray}
&&
D_1=k^2,
D_2=l^2,
D_3=(p+k)^2-m^2,
D_4=(p+l)^2-m^2,
\nonumber\\&&
D_5=(p+k+l)^2-m^2,
D_6=p^2-m^2
\,.
\nonumber
\end{eqnarray}
The coefficient functions from Eq.~(\ref{solution}) read:
\begin{eqnarray}
f(\underline{n},D)&=&
\Df{1}{(2\pi\imath)^6}
\oint\Df{dx_1}{x_1^{n_1}}
\cdots
\oint\Df{dx_6}{x_6^{n_6}}
{P(x_1,x_2,x_3+1,\dots,x_6+1)^{D/2-2}}
\,,
\label{solution3}
\end{eqnarray}
where the polynomial $P(x_1,\ldots,x_6)$ is given by
\begin{eqnarray}
P(x_1,\dots,x_6)&=&
(x_1+x_2)(x_1x_2-x_3x_4-x_5x_6)+(x_3+x_4)(-x_1x_2+x_3x_4-x_5x_6)
\nonumber\\&&\mbox{}
+(x_5+x_6)(-x_1x_2-x_3x_4+x_5x_6)
+x_1x_3x_6+x_1x_4x_5+x_2x_3x_5
\nonumber\\&&\mbox{}
+x_2x_4x_6
\,.
\nonumber
\label{eqpx}
\end{eqnarray}
In this equation we have omitted an overall factor $(-1/4)$
as it would lead to a trivial rescaling which cancels after considering 
initial conditions.

Let us now compute the coefficient functions 
which correspond to the choice of the master integrals given 
in~\cite{REC,Avd}
where $B(\underline{n},D)$ has been written in the form:
\begin{eqnarray}
B(\underline{n},D)&=&
N(\underline{n},D)B(0,0,1,1,1,1,D)+
M(\underline{n},D)B(1,1,0,0,1,1,D)
\nonumber\\&&\mbox{}
+
T(\underline{n},D)B(0,0,0,1,1,1,D)
\,.
\nonumber
\end{eqnarray}
This leads us to the following normalization conditions:
\begin{eqnarray}
N(0,0,1,1,1,1,D)=1,&\quad N(1,1,0,0,1,1,D)=0,&\quad 
N(0,0,0,1,1,1,D)=0,\label{condN}\\
M(0,0,1,1,1,1,D)=0,&\quad M(1,1,0,0,1,1,D)=1,&\quad
M(0,0,0,1,1,1,D)=0,\nonumber%
\label{condM}
\\
T(0,0,1,1,1,1,D)=0,&\quad T(1,1,0,0,1,1,D)=0,&\quad 
T(0,0,0,1,1,1,D)=1
\,.
\label{condT} 
\end{eqnarray}
In a first step the function $N(\underline{n},D)$ is considered.
The last two conditions of Eq.~(\ref{condN}) are satisfied if 
the contours for the massive 
indices are chosen to be small circles around zero, i.e., $x_i=0$. 
In practice it is useful to perform a Taylor expansion in 
$x_3, x_4, x_5$ and $x_6$ where, according to the
residuum theorem, only one term leads to a result different from
zero (Here and in the following the proportional sign (``$\propto$'')
is used as the $D$ dependent factor is fixed at the end in accordance
with the normalization given in Eq.~(\ref{condN}).):
\begin{eqnarray}
N(\underline{n},D)&\propto&\oint\oint
\Df{dx_1dx_2}{x_1^{n_1}x_2^{n_2}}
\left[\Df{\partial_3^{n_3-1}\dots\partial_6^{n_6-1}}
{(n_3-1)!\dots(n_6-1)!}
P(x_1,x_2,x_3+1,\dots,x_6+1)^{D/2-2}
\right]
\bigg|_{x_3,\dots,x_6=0}.\nonumber
\end{eqnarray}
The remaining integrals over $x_1$ and $x_2$, for which in general
(because of the derivatives)
the exponent of the polynomial $P(x_1,\ldots,x_6)$ is reduced,
can easily be 
expressed in terms of Pochhammer symbols $(a)_n = \Gamma(a+n)/\Gamma(a)$:
\begin{eqnarray}
\oint\oint
\Df{dx_1dx_2}{x_1^{n_1}x_2^{n_2}}
P(x_1,x_2,1,1,1,1)^{D/2-2-c}
\propto 
\Df{(D/2-1)_{-c}(4-3D/2)_{n_1+n_2+3c}}
      {4^{(n_1+n_2+3c)}(2-D/2)_{n_1+c}(2-D/2)_{n_2+c}},
\nonumber
\end{eqnarray}
where $\Gamma(x)$ is Eulers $\Gamma$ function.

Let us next turn to the function $M(\underline{n},D)$ which can be
treated in analogy. The only difference is that 
due to the symmetry $B(1,1,0,0,1,1,D)=B(1,1,1,1,0,0,D)$ it is profitable
to consider the sum of the solutions 
with a Taylor expansion performed in the variables
$(x_1,x_2,x_3,x_4)$ and $(x_1,x_2,x_5,x_6)$:
\begin{eqnarray}
M(\underline{n},D)&=&
m(n_1,n_2,n_3,n_4,n_5,n_6,D)+m(n_1,n_2,n_5,n_6,n_3,n_4,D)
\,,
\nonumber
\end{eqnarray}
where $m(\underline{n},D)$ is given by
\begin{eqnarray}
m(\underline{n},D)&\propto&
\oint\oint
\Df{dx_5dx_6}{x_5^{n_5}x_6^{n_6}}
\left[\Df{\partial_1^{n_1-1}\dots\partial_4^{n_4-1}}
{(n_1-1)!\dots(n_4-1)!}
P(x_1,x_2,x_3+1,\dots,x_6+1)^{D/2-2}
\right]
\bigg|_{x_1,\dots,x_4=0}
\,.
\nonumber
\end{eqnarray}
Again the remaining integrals are easily performed with the result
\begin{eqnarray}
\lefteqn{
\oint\oint
\Df{dx_5dx_6}{x_5^{n_5}x_6^{n_6}}
P(0,0,1,1,x_5+1,x_6+1)^{D/2-2-c}
}
\nonumber\\
&\qquad\qquad\qquad\qquad\qquad\qquad
\propto&
\Df{(-)^{n_5+n_6}(3-D)_{n_5+2c}(3-D)_{n_6+2c}
(4-3D/2)_{n_5+n_6+3c}}{(2-D/2)_c(6-2D)_{n_5+n_6+4c}(3-D)_{n_5+n_6+2c}}
\,.
\nonumber
\end{eqnarray}

The case ``T'' is more complicated. The conditions~(\ref{condT}) 
show that we should find a solution which is non-zero
if one of the massive indices is non-positive. It turns out
that the following combination of ``Taylor'' solutions
leads to the desired result:
\begin{eqnarray}
T(n_1,n_2,n_3,n_4,n_5,n_6,D)&=&
t(n_1,n_2,n_3,n_4,n_5,n_6,D)+t(n_1,n_2,n_4,n_3,n_6,n_5,D)
\nonumber\\
&&\mbox{}
+t(n_1,n_2,n_5,n_6,n_3,n_4,D)+t(n_1,n_2,n_6,n_5,n_4,n_3,D),
\nonumber
\end{eqnarray}
where
\begin{eqnarray}
t(\underline{n},D)=0\ \quad\mbox{if}\quad\ 
       n_4\ \mbox{or}\ n_5\ \mbox{or}\ n_6\ <\ 1.
\label{t3}
\end{eqnarray}
Four terms are necessary in order to meet the symmetry properties of the
initial integral. From Eq.~(\ref{solution3}) the following general 
representation for the solution with the property given in Eq.~(\ref{t3})
is obtained:
\begin{eqnarray}
\overline{t}(\underline{n},D)&\propto&\oint\oint
\Df{dx_1dx_2dx_3}{x_1^{n_1}x_2^{n_2}x_3^{n_3}}
\nonumber\\&&\mbox{}
\quad\times
\left[\Df{\partial_4^{n_4-1}\partial_5^{n_5-1}\partial_6^{n_6-1}}
{(n_4-1)!(n_5-1)!(n_6-1)!}
P(x_1,x_2,x_3+1,\dots,x_6+1)^{D/2-2}
\right]
\bigg|_{x_4,x_5,x_6=0}
\,,
\label{t31}
\end{eqnarray}
where we have introduced the function $\overline{t}(\underline{n},D)$
as in general the expression given on the r.h.s. of Eq.~(\ref{t31})
leads to a mixture of the functions $N(\underline{n},D)$ and
$t(\underline{n},D)$.
We will moreover see below that $\overline{t}(\underline{n},D)$
obeys simple recurrence relations and in addition the extraction
of $t(\underline{n},D)$ and $N(\underline{n},D)$ is quite simple.
The resulting integral where $x_4=x_5=x_6=0$ is given by:
\begin{eqnarray}
\overline{t}(n_1,n_2,n_3,D)&\equiv&
\overline{t}(n_1,n_2,n_3,1,1,1,D)
\nonumber\\&=&
\Df{1}{(2\pi\imath)^3}
\oint\oint\oint
\Df
{dx_1dx_2dx_3}
{x_1^{n_1}x_2^{n_2}x_3^{n_3}}
P(x_1,x_2,x_3+1,1,1,1)^{D/2-2}
\nonumber
\\&=&
\Df{1}{(2\pi\imath)^3}
\oint\oint\oint
\Df
{dx_1dx_2dx_3}
{x_1^{n_1}x_2^{n_2}x_3^{n_3}}
{(x_3^2-x_1x_2x_3+x_1x_2(x_1+x_2-4))^{D/2-2}}
\,.
\label{tbar}
\end{eqnarray}
The function $\overline{t}(n_1,n_2,n_3,D)$ obeys the 
following recurrence relations
which can be derived by using integration-by-parts in Eq.~(\ref{tbar}):
\begin{eqnarray}
{\bf I}^+_1 \overline{t}(n_1,n_2,n_3,D+2)&=&
(D/2-1)\left[-{\bf I}^-_2{\bf I}^-_3 + 2{\bf I}^-_1{\bf I}^-_2 
+ ({\bf I}^-_2)^2 -4 {\bf I}^-_2\right]\overline{t}(n_1,n_2,n_3,D),
\nonumber\\&&
\label{rn1}\\
{\bf I}^+_2 \overline{t}(n_1,n_2,n_3,D+2)&=&
(D/2-1)\left[-{\bf I}^-_1{\bf I}^-_3 + 2{\bf I}^-_1{\bf I}^-_2 
+ ({\bf I}^-_1)^2 -4 {\bf I}^-_1\right]\overline{t}(n_1,n_2,n_3,D),
\nonumber\\&&
\\
{\bf I}^+_3 \overline{t}(n_1,n_2,n_3,D+2)&=&
(D/2-1)\left[2{\bf I}^-_3 - {\bf I}^-_1{\bf
I}^-_2\right]\overline{t}(n_1,n_2,n_3,D),
\label{eqrdrd}
\\
\overline{t}(n_1,n_2,n_3,D+2)&=&
\left[({\bf I}^-_3)^2 
+{\bf I}^-_1{\bf I}^-_2(-{\bf I}^-_3 + {\bf I}^-_1 + {\bf I}^-_2 -4)\right]
\overline{t}(n_1,n_2,n_3,D)\label{rd}.
\end{eqnarray}
The last equation immediately follows from~(\ref{rrD0}).

For $n_3<1$ the contribution from $N(\underline{n},D)$ 
vanishes as the integration around
$x_3=0$ gives zero and only $t(\underline{n},D)$ survives.
In this case with the simple change of variables,
$x_3=\sqrt{y_3}+x_1x_2/2$, the integral~(\ref{tbar}) can be reduced to a
sum over integrals which furthermore can be
expressed in terms of Pochhammer symbols.
Keeping in mind the normalization condition~(\ref{condT}) 
we get the desired solution for $t(\underline{n},D)$:
\begin{eqnarray}
t(n_1,n_2,n_3<1,D-2c)
&=&\Df{\overline{t}(n_1,n_2,n_3,D-2c)}{\overline{t}(0,0,0,D)}
\nonumber\\
&=&\Df{(D/2-1)_{-c}(D/2-1/2)_{-c}}{(-)^{n_3+c}\,2^{(2n_1+2n_2+6c+3n_3)}}
\Df
{(2-D)_{(n_1+n_3+2c)}(2-D)_{(n_2+n_3+2c)}}
{(3/2-D/2)_{(n_1+n_3+c)}
(3/2-D/2)_{(n_2+n_3+c)}}
\nonumber\\&&\mbox{}
\times\sum_{k=0}^{[-n_3/2]}
\Df{
(1/2)_k
(n_3)_{(-n_3-2k)}
(D/2-1/2-c)_k
}{
(-n_3-2k)!
(3/2-D/2+n_1+n_3+c)_k(3/2-D/2+n_2+n_3+c)_k
}.
\label{n30}
\end{eqnarray}

For $n_3>1$ one can reduce $n_3$ to 1 with the help of:
\begin{eqnarray}
\overline{t}(n_1,n_2,n_3>1,D) & \propto &
\oint\oint\oint
\Df
{dx_1dx_2dx_3}
{x_1^{n_1}x_2^{n_2}x_3}
\nonumber\\&&\mbox{}\qquad\times
\left[
\Df{\partial_3^{n_3-1}}
{(n_3-1)!}
{(x_3^2-x_1x_2x_3+x_1x_2(x_1+x_2-4))^{D/2-2}}
\right]
\,.
\label{tbar1}
\end{eqnarray}
In connection with the reduction of $n_3$ two remarks are in order:
First, we should mention that the reduction of $n_3>1$ to
$n_3=1$ can also be expressed through a finite sum which is obvious form
Eq.~(\ref{tbar1}).
Second, also for $n_3<1$ the explicit sum in Eq.~(\ref{n30}) can be
replaced by a recurrence procedure using Eqs.~(\ref{eqrdrd}) and
(\ref{rd}) in order to arrive at $n_3=0$. Afterwards
$n_1$ and $n_2$ are reduced to zero using
Eq.~(\ref{n30}) with $n_3=0$.

For the case $n_3>1$ one gets after the reduction to $n_3=1$
a set of functions
$\overline{t}(n_1,n_2,1,D-2c)$ with various $n_1, n_2$ and $c$. 
The further reduction is done with the help of the relations:
\begin{eqnarray}
\overline{t}(n_1,n_2,1,D)&=&\Df{(D-4)}{(2n_1-D+2)}
\left[\overline{t}(n_1-2,n_2-1,1,D-2)
-\Df{1}{2}\overline{t}(n_1-1,n_2-1,0,D-2)\right],
\nonumber\\&&
\label{rt1}\\
\overline{t}(n_1,n_2,1,D)&=&
\Df{(2n_2-D+4)}{(2n_1-D+2)}
\overline{t}(n_1-1,n_2+1,1,D)
+\Df{(n_1-n_2-1)}{(2n_1-D+2)}
\overline{t}(n_1,n_2+1,0,D)
\,,
\label{rt2}
\end{eqnarray}
which can be obtained from Eqs.~(\ref{rn1})-(\ref{rd}).
Note that $\overline{t}(n_1,n_2,1,D)=\overline{t}(n_2,n_1,1,D)$.
Therefore these equations can be used to reduce $(n_1,n_2)$
to $(0,-1)$, $(1,0)$, $(0,0)$ and further combinations where
$n_3=0$. Then, using Eqs.~(\ref{rn1}) and~(\ref{rd}) 
for $n_1,n_2=-1,0,1,2$, respectively, we get:
\begin{eqnarray}
\overline{t}(0,-1,1,D)&=&\Df{4}{3}\overline{t}(0,0,1,D)
+\Df{1}{3}\overline{t}(0,0,0,D),\label{s1}\\
\overline{t}(1,0,1,D)&=&\Df{(3D-8)}{4(D-4)}\overline{t}(0,0,1,D)
-\Df{(D-2)^2}{8(D-3)(D-4)}\overline{t}(0,0,0,D)
\,.\label{s2}
\end{eqnarray}
Finally the shift in the last argument is fixed with the help of
\begin{eqnarray}
\overline{t}(0,0,1,D+2)&=&
-\Df{2(D-2)}{3(3D-2)(3D-4)}\left[32(D-2)\overline{t}(0,0,1,D)
+(11D-16)\overline{t}(0,0,0,D)\right]
\,,
\nonumber\\&&
\label{s3}
\\
\overline{t}(0,0,0,D-2c)&=&
\Df{(-)^c(D/2-1)}{4^c(D/2-1-c)}
\Df{(D/2-1/2)_{-c}}{(D/2)_{-c}} 
\,
\overline{t}(0,0,0,D)
\label{s4}
\,,
\end{eqnarray}
which completes the reduction of $\overline{t}(n_1,n_2,n_3,D-2c)$ to
$\overline{t}(0,0,1,D)$ and $\overline{t}(0,0,0,D)$. 
The result for $t(n_1,n_2,n_3,D-2c)$ is obtained from:
\begin{eqnarray}
t(n_1,n_2,1,D) &=& \frac{1}{\overline{t}(0,0,0,D)}
\left[
\overline{t}(n_1,n_2,1,D)
-\overline{t}(0,0,1,D) N(n_1,n_2,1,1,1,1)
\right]
\,,
\end{eqnarray}
which follows from Eq.~(\ref{condT}). This means that in practice 
from the final expression of the reduction procedure 
simply the coefficient of 
$\overline{t}(0,0,0,D)$ has to be extracted in order to get
the result for $t(\underline{n},D)$.
Note that due to Eq.~(\ref{n30}) the functions
$\overline{t}(n_1,n_2,0,D-2c)$ in 
Eqs.~(\ref{rt1})-(\ref{s3})
are rational in $D$, i.e., the recursion for $n_1, n_2$ and $c$
are simple one-parameter equations of the type $F(n)=a(n)F(n-1)+b(n)$ 
where $a(n)$ and $b(n)$ are known rational functions. 
Thus they can easily be solved and expressed through 
sums of Pochhammer symbols. At this point we refrain from listing
the resulting formulas explicitly.

\section{\label{secimp}Implementation}

In this section a possible implementation of the above formulas
is discussed.
We will mainly speak about the most difficult coefficient function (``T''),
the others (``M'' and ``N'') constitute simple sub-cases and can be 
considered in complete analogy.
The calculation of the ``T'' coefficient function implies subsequent 
reduction of:\\
1) $n_4, n_5, n_6$ to 1,\\
2) $n_3$ to 0 or 1,\\
3) $n_1, n_2$ to (0,0), (1,0), (0,-1),\\
4) $D-2c$ to $D$.\\
The steps 1) and 2) demand a Taylor expansion of a polynomial raised to 
non-integer power. This can either be done by direct differentiation
(see Eqs.~(\ref{t31}) and (\ref{tbar1})), or by computing multiple
combinatoric sums which arise after rewriting the action
of the operators $\partial^n/\partial x_i^n$.
Moreover, the reduction of each parameter can be done 
independently by one of these methods.
The steps 3) and 4) can be performed either by using the 
recursion relations (\ref{rt1})-(\ref{s4})
or by explicitly evaluating the sums.

In practice, the quantities of physical interest can be expressed as
linear combination of vacuum integrals, the number of which can be
very large (up to $10^7$; see, e.g.~\cite{chetn}). It is very natural 
to perform the
reductions 1)-4) not for individual integrals, but for the whole
expression as simplifications may take place in intermediate steps of the
calculation.
Of course, the method which gives the best performance strongly depends
on the computer algebra system used. In the following we 
compare REDUCE~3.6 and FORM~2.3 both running on a DEC-Alpha-station
with $266$~MHz.

\vspace{.5em}

REDUCE.
Our experience shows that the method using differentiation 
in steps 1) and 2) works about two times faster, although the 
calculation of the multiple sums allows to save computer memory.

At step 3) the recurrence procedure is also about two times faster
than the use of explicit sums. The higher demand concerning the memory 
is not essential at this point as 
for step 1) more memory is needed and thus the limitations 
are given from there.
Step 4) needs negligible time and memory. Here in practice the
recursion approach is used.

\vspace{.5em}

FORM.
The explicit differentiation using FORM requires
for the variables to be defined as ``non-commuting''~\cite{FORM}.
This has the consequence that the sorting of the intermediate
expressions becomes quite slow and needs in addition more main memory.
Therefore in practice only the method relying on multiple sums 
are effective for step 1). At steps 2) and 3) we can choose between
the combinatoric-sum and the recursion approach. It turned out that
their performance is comparable. At step 4) we prefer to use a
one-parameter table for the functions $t(0,0,1,D-2c)$.

The implementation in FORM is done in such a way that
an expansion in $D-4$ is performed for the intermediate
expressions. Thereby some tricks were used like the
application of the ``ACCU'' function which collects polynomials in $D-4$
or the use of tables for the Pochhammer symbols~\cite{FORM}.
Note that explicit formulas are used for the
realization. Therefore one has full control on the powers of 
intermediate poles in $D-4$. This actually is a problem in the
approach based on recurrence relations only. There the expansion in
$D-4$ has to be performed to higher order which has to be fixed
after careful examination of the recurrence relations.

\vspace{.5em}

It is interesting to mention that the implementation within REDUCE,
which even computes the full $D$-dependence,
is about three times faster than the one using FORM.
An explanation for this unexpected fact
is the possibility to use 
direct differentiation within REDUCE.
It might also be possible that during the calculation
essential cancellations among long sums containing Pochhammer symbols 
take place. REDUCE with its ability to work with rationals can 
exploit this fact, however, FORM cannot.
We should also mention that the attempt to expand in $D-4$ in the
``REDUCE'' version (using ``wtlevel'' declaration~\cite{REDUCE}) takes more
time but is still faster than FORM.

Within FORM we are also in the position to compare the recursion 
approach~\cite{ch-tk,REC} where, however, in addition to the
recurrence relations also tables were used~\cite{MATAD},
with the one described in this paper.
As an example we considered the contribution of the non-planar diagram to the
photon propagator and performed an expansion up to ${\cal O}(q^{12})$
where $q$ is the external momentum. It turned out, that the ``improved'' 
recursion works better for planar vacuum integrals contributing to this
diagram, however, worse for the non-planar ones. The joined variant which
uses ``improved'' recursion for the planar diagrams and the 
new method for non-planar ones gives  a gain of roughly $8$~\% as compared to 
the use of only the ``improved''  recursion.

\section{\label{seccon}Conclusions}

The new approach described in this paper allows to solve the
recurrence relations for three-loop vacuum integrals explicitly
and expresses the result in terms of multiple 
sums containing Pochhammer symbols.
Nevertheless, the final choice of the algorithm which leads to 
the best performance
strongly depends on the peculiarities of the problem and also
on the symbolic language used.

For the particular problem considered here --- three-loop integrals
with four massive and two massless lines --- we get better results using
REDUCE~3.6 than using FORM~2.3 from point of view of CPU time and
in the presence of sufficient memory resources. 
Certainly, FORM stands beyond
the competition as soon as the intermediate expressions become
very large and enormous swapping to the hard disk is necessary.

\section*{Acknowledgment}

We would like to thank J.H.~K\"uhn and K.G.~Chetyrkin for stimulating
discussions. P.A.B. thanks the University of Karlsruhe for the hospitality 
during the visit when an important part of this work has been done.


\begin{thebibliography}{99}

\bibitem{BaiBro95}
P.\,A.\,Baikov and D.\,J.\,Broadhurst,
Presented at 4th International Workshop on Software Engineering and
Artificial Intelligence for High Energy and Nuclear Physics (AIHENP95), 
Pisa, Italy, 3-8 April 1995. 
Published in Pisa AIHENP (1995) 167.

\bibitem{chetn} %
K.\,G.\,Chetyrkin, J.\,H.\,K\"uhn and M.\,Steinhauser,  
{Phys. Lett.} {\bf B371} (1996) 93
{Nucl. Phys.} {\bf B482} (1996) 213;
{Nucl. Phys.} {\bf B505} (1997) 40.

\bibitem{ch-tk} %
K.\,G.\,Chetyrkin and F.\,V.\,Tkachov, Nucl.\,Phys.\,{\bf B192} (1981) 159;\\
F.\,V.\,Tkachov, Phys.\,Lett.\,{\bf B100} (1981) 65.

\bibitem{REC} %
D.\,J.\,Broadhurst, Z.\,Phys.\,{\bf C54} (1992) 559.

\bibitem{Tar97} %
O.\,V.\,Tarasov, Nucl.\,Phys.\,{\bf B502} (1997) 455.

\bibitem{Avd} %
L.\,V.\,Avdeev, 
Comput.\,Phys.\,Commun. {\bf 98} (1996) 15.

\bibitem{Tka83}
F.\,V.\,Tkachov, Theor. Mat. Fiz. {\bf 56} (1983) 350.

\bibitem{mincer} 
S.\,A.\,Larin, F.\,V.\,Tkachev and J.\,A.\,M.\,Vermaseren,
NIKHEF Report No.\ NIKHEF--H/91--18 (September 1991).

\bibitem{ES} %
P.\,A.\,Baikov, Phys.\,Lett.\,{\bf B385}, (1996) 404;
Nucl.~Instr.~\& Methods~{\bf A389} (1997) 347.

\bibitem{REDUCE} %
A.\,C.\,Hearn, REDUCE User's Manual, version 3.6.

\bibitem{FORM} %
J.\,A.\,M.\,Vermaseren, Symbolic Manipulation with FORM, version 2.

\bibitem{MATAD}
M. Steinhauser, Ph.D. thesis, Karlsruhe University
(Shaker Verlag, Aachen, 1996).


\end{thebibliography}
\end{document}